\documentclass[12pt,preprint]{aastex}
\usepackage{psfig,emulateapj5}

\newcommand{\kms}{km~s$^{-1}$}

\newcommand{\msun}{M$_{\odot}$}

\begin{document}

\slugcomment{}

\title{VLA Imaging of the Intriguing HI Cloud HIJASS J1021+6842 
in the M\,81 Group}

\author{Fabian Walter}
\affil{Max-Planck-Institute for Astronomy, K\"onigstuhl 17,
  Heidelberg, D-69117, Germany} 
\email{walter@mpia-hd.mpg.de}
\author{Evan D.\ Skillman}
\affil{Astronomy Department, University of Minnesota, 116 Church St. SE,}
\affil{Minneapolis, MN 55455, USA}
\email{skillman@astro.umn.edu}
\and
\vspace{-5mm}
\author{Elias Brinks}
\affil{Centre for Astrophysics Research, University of
Hertfordshire, College Lane, Hatfield, Hertfordshire, AL10 9AB, UK}
\affil{and INAOE, Apdo.\ Postal 51, Puebla, Pue 72000, Mexico}
\email{ebrinks@star.herts.ac.uk}

\begin{abstract}
  We present VLA HI 21cm observations of HIJASS J1021+6842 which has
  been discovered in the direction of the M\,81 group.  Our synthesis
  imaging reveals that the HI is distributed over a larger angular
  extent and velocity range than the single dish discovery
  observations. Assuming that HIJASS J1021+6842 is at the distance of
  the M\,81 group, we detect 1.5 $\times$ 10$^8$ \msun\ of HI
  distributed over as much as 30 kpc, i.e., substantially larger than
  the biggest dwarf galaxies in the same group. At the depth of our
  imaging, the HI appears to be confined to at least 7 clouds.  Peak
  HI column densities are $\sim1.8 \times 10^{20}$ atoms cm$^{-2}$
  which is well below the canonical star formation threshold of
  $\sim10^{21}$ atoms cm$^{-2}$ and therefore consistent with the fact
  that no optical counterpart has as yet been identified.  A gradient
  in velocity is observed across the extent of the detected HI;
  assuming that the object is gravitationally bound we derive a
  dynamical mass of $7\times10^9$\,M$_{\odot}$ and a
  dark--to--luminous mass ratio of $>$10. Alternatively, a tidal
  origin may also result in the observed velocity gradient which would
  lead to a considerably lower dynamical mass. Given the above
  properties and the absence of evidence of a stellar population,
  HIJASS J1021+6842 is unique amongst the other systems in the M81
  group.
\end{abstract}


\section{Introduction}

The nearby M\,81 group provides an ideal
testbed for studies of dwarf galaxies and the effects of interactions
between galaxies. Thanks to its proximity, it is one of the best
studied groups, both regarding the interaction between its
three luminous galaxies \citep[M\,81, M\,82 and NGC\,3077,
e.g.,][]{YHL94}, and the studies of individual group members, notably
several of the dwarf galaxies \citep[e.g.,][] {PWBR92, WB99, WWMS02,
  Oea01}.

\citet{Bea01} recently conducted a blind HI survey of the M\,81 group
using a multibeam receiver on the 76--m Lovell telescope with a
resultant beamsize of $\approx$ 12\arcmin .  Interestingly, only one
new object was discovered in their observations, HIJASS J1021+6842,
which is the subject of this letter.  This implies that the census of
galaxies with HI gas in the M\,81 group must be nearly complete down
to HI masses of order 10$^7$ \msun.

HIJASS J1021+6842 lies at an angular distance of $105^\prime$ (or a
minimum separation of 110 kpc at the distance of M\,81) from IC\,2574.
The proximity in sky position and radial velocity between IC\,2574 and
HIJASS J1021+6842 implies that HIJASS J1021+6842 is a probable member
of the M\,81 group at a distance of $\sim4$\,Mpc and a possible
companion of IC\,2574 at a distance of $\sim4$\,Mpc
(\citet{Bea01, Kea02}). \citet{Bea01} noted the lack of an optical
counterpart for HIJASS J1021+6842 in the second--generation red
Digital Sky Survey, and speculated that it may be either a very low
surface brightness companion of IC\,2574, or the debris of a tidal
encounter between IC\,2574 and one of the galaxies around M\,81.

Thus, HIJASS J1021+6842 is an intrinsically interesting object, and
potentially a key object for understanding the process of galaxy
formation from tidal debris.  The present higher resolution HI
observations provide a requisite first step to understanding the true
nature of this object.

\section{Observations and Analysis}

\begin{figure}
\epsscale{1}
\plotone{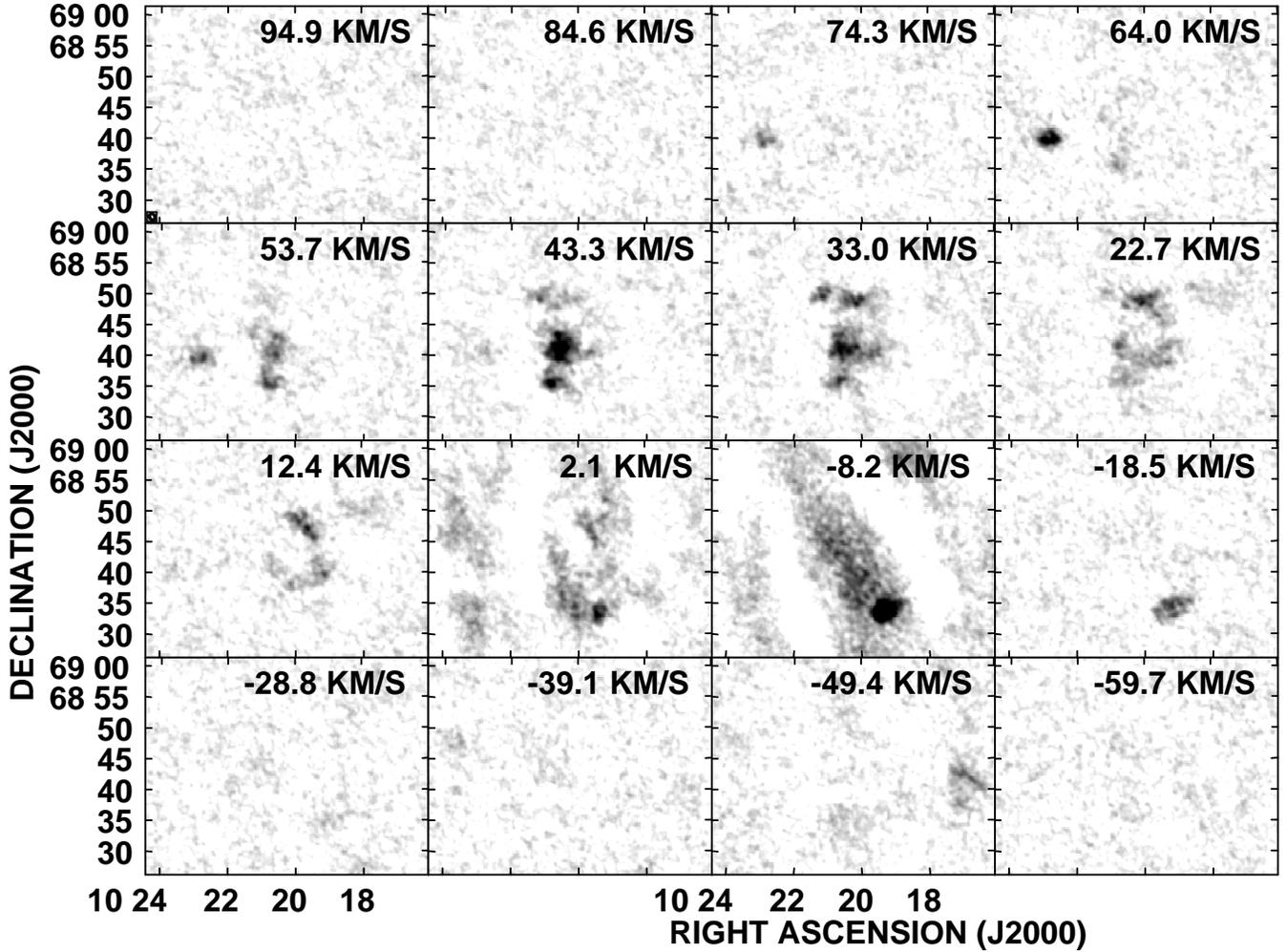}
\caption{HI channel maps of HIJASS J1021+6842 (every other  channel is 
  shown, primary beam attenuation has not been applied). The greyscale
  displays a range from 0 to 9\,mJy\,beam$^{-1}$.  The beamsize
  ($60''\times52''$) is indicated in the lower left of the top left
  panel.  Note that the channel maps at around 0 km\,s$^{-1}$ are
  affected by large scale Galactic HI emission.  }
\label{mosaic}
\end{figure}

We used the NRAO\footnote{The National Radio Astronomy Observatory is
  a facility of the National Science Foundation operated under
  cooperative agreement by Associated Universities, Inc.}  Very Large
Array (VLA) to obtain HI spectral line observations of HIJASS
J1021+6842 in the D configuration (integration time: 106 minutes) and
in the C configuration (148 minutes) in 2004. We observed
0841+708 as the complex gain calibrator and 0542+498 (3C147) and
1331+305 (3C286) as flux and bandpass calibrators. The spectra
consisted of 128 channels with a width of 5.2~km\,s$^{-1}$ (after
on--line Hanning smoothing). The data were analyzed using standard
calibration/mapping tasks in the AIPS software package.  Four
  channels near velocities of 0~km\,s$^{-1}$ were contaminated by
  Galactic foreground emission (see Fig.~\ref{mosaic}).  To calculate
  the integrated HI map (moment 0) we carefully inspected and blanked
  consecutive channel maps to remove the extended Galactic emission
  from the data cube. In doing so we are helped by the fact that
  smoothly distributed emission, such as that due to foreground
  Galactic HI, is filtered out in interferometric observations.

Given the intrinsically faint HI emission of HIJASS J1021+6842, images
were made with a uvtaper of 5.3 k$\lambda$ to emphasize weak, extended
structures. This resulted in a beamsize of $60''\times52''$
(60$''$=1.17\,kpc) and an rms of 0.8\,mJy\,beam$^{-1}$ per channel.
Given the considerable extent of the HI emission, it was important to
correct for primary beam attenuation in all subsequent, quantitative
analysis.  \vspace{-3mm}

\section{HI Distribution and Kinematics}

\subsection{HI Distribution}

Figure~\ref{mosaic} shows a mosaic of 12 channel maps (before primary
beam correction) stepped at every other channel.  This mosaic shows
that HI is detected from $\sim$ $-$50 to $\sim$$+$70 km\,s$^{-1}$ in
radial velocity, which is a larger range in velocity than the FWHM of
$\approx$ 50 km\,s$^{-1}$ reported by \citet{Bea01}. It is likely that
the synthesis observations have allowed us to overcome confusion with
Galactic HI emission, enabling us to detect HI at velocities near
0~km\,s$^{-1}$ and lower velocities.  Additionally, the higher
resolution imaging shows that the single source discovered by
\citet{Bea01} breaks up into several regions of emission.  There is a
radial velocity gradient with positive velocities in the eastern
components and negative velocities in the western components.  What is
particularly striking about these images is the fact that the HI is
extended over more than 30\,kpc which makes HIJASS J1021+6842 bigger
than even the most massive dwarf galaxies in the M\,81 group (e.g.,
IC\,2574: \citet{WB99} Ho\,II: \citet{PWBR92}).

\begin{figure}
\epsscale{1}
\plotone{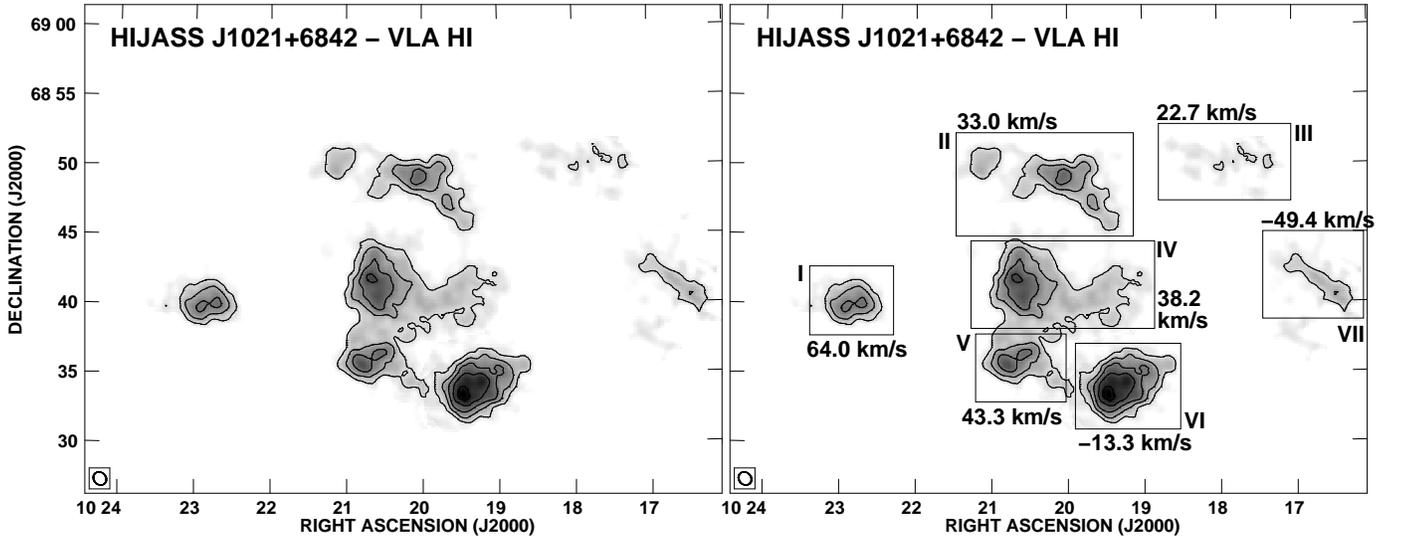}
\caption{The total HI column density image of HIJASS J1021+6842 
  (corrected for primary beam attenuation). The field size is
  identical to that of Figure 1.  The beamsize ($60''\times 52''$) is
  indicated in the lower left.  Contours are plotted at 4, 7, 10, 13,
  16$\times$10$^{19}$\,atoms\,cm$^{-2}$. The total HI mass is
  1.5$\times$10$^{8}$\,M$_{\odot}$.  The boxes in the right hand panel
  indicate regions with coherent structures in position-velocity
  space. Masses for the individual regions are (in units of
  10$^{7}$\,M$_{\odot}$): I: 1.3, II: 2.6, III: 0.8, IV: 4.1, V: 1.5,
  VI: 3.0, VII: 1.1. The numbers adjacent to the boxes give the
  velocities at which the individual regions appear brightest in the
  channel maps. Note an apparent gradient in velocity ranging from
  around 70\,\kms (region I) to -50\,\kms (region VII).  }
\label{mom0}
\end{figure}

Figure~\ref{mom0} shows a map of the total HI column density of HIJASS
J1021+6842. We identify seven regions of emission (labeled I--VII in
Fig.~\ref{mom0}).  The maximum HI column density is found in region VI
and corresponds to $1.8 \times 10^{20}$ atoms cm$^{-2}$. Region
III is a marginal detection and needs follow up observations.

The total HI emission shown in Fig.~\ref{mom0} corresponds to a total
HI mass of 1.5 $\times$ 10$^8$ \msun .  This is considerably larger
than the value reported by \citet{Bea01}.  We attribute the difference
to the fact that most of the emission is situated in small clumps
(compared to the Lovell beam, FWHM: $\sim$12$'$), that the area over
which the clumps are spread out is larger than the size of the Lovell
beam, and that HIJASS J1021+6842 extends across velocities containing
Galactic HI emission.  The HI masses of the individual regions
are listed in the caption of Fig.\,\ref{mom0}.

\subsection{HI Kinematics and Dynamical Mass Estimate}

The channel maps (Fig.~\ref{mosaic}) show that there is a velocity
gradient across the HI extent of HIJASS J1021+6842 extending from
region I (near 70 km\,s$^{-1}$) to region VII ($-$50 km\,s$^{-1}$).
The velocities at which the individual regions appear brightest in the
channel maps are given in Fig.~\ref{mom0} (right).

One interpretation of the velocity gradient is that the emission is
gravitationally bound and that the velocity gradient can be
interpreted as being due to a (broken) annulus in rotation. Under this
assumption we find for a radius of 15 kpc and a V$_{max}$ of 40
km\,s$^{-1}$ a minimum dynamical mass of $5.5 \times 10^9$ \msun.  The
resulting ratio of dynamical over detected mass (or M$_{\rm
  dyn}$/M$_{\rm HI}$) is $>$ 10 for the ensemble. This would be a
large ratio which is higher than is typically observed in lower
luminosity dwarf irregular galaxies \citep{S96}. We stress however,
that a tidal origin may also result in the observed velocity gradient
which would imply a significantly lower dynamical mass.
\vspace{-3mm}
\section{Constraints on the True Nature of HIJASS J1021+6842}

\subsection{Lack of an Optical Counterpart}

As noted by \citet{Bea01}, there is no obvious optical counterpart to
HIJASS J1021+6842 which is detected on images of the DSS. In addition
to this, the M\,81 group has been the subject of many very thorough
searches for small and low surface brightness members
\citep[e.g.,][]{vdb59, K68, BK82, Cea98, KK98}.  These searches
reached faint surface brightness levels and lead to the discovery of
dwarfs as faint as m$_{\rm V}\sim18$, corresponding to M$_{\rm
  V}\sim-10$ at the distance of the M81 group (thus reaching down to
the extreme low--mass end of the luminosity function). As yet, none of
these searches have come up with candidate counterparts for HIJASS
J1021+6842 \citep[e.g.,][]{Cea98}. One possible concern is the presence of
a bright star in the field (HD 89343, RA(2000.0) = 10 21 03.3,
DEC(2000.0) = $+$68 44 52, V = 6 mag.)  which may limit the local
sensitivity to low surface brightness optical emission.

It should be noted, though, that the HI distribution is so different
from the other M\,81 group dwarfs that perhaps one would not
necessarily expect stars to be associated with this system.  Normally,
star formation is associated with HI column densities above a certain
threshold \citep[e.g.,][]{K89}.  Empirically this threshold hovers
around $10^{21}$ atoms cm$^{-2}$ for dwarf irregular galaxies. The
fact that no optical counterpart has been found appears to be
consistent with this threshold.  Clearly, deep optical follow--up
observations are needed to determine at higher confidence whether or
not stars are associated with HIJASS J1021+6842.

Another possible explanation for the lack of stars in HIJASS
J1021+6842 is the possibility that we are seeing the early stages of
formation of a new galaxy.  This could either be formation from a
primodial cloud or formation from tidal debris as discussed, e.g., by
\citet{Mea02} or \citet{Bea04} (and references therein).  
Given the relatively dense environment of the central M\,81 triplet,
the position of HIJASS J1021+6842 seems to be an unlikely location for
the contemporary precipitation of a primordial galaxy.  If HIJASS
J1021+6842 is tidal debris (perhaps from the outer parts of one of the
larger galaxies in the M\,81 group), then the small velocity
differences between its components allow for the possibility that the
entire system is gravitationally bound, which could potentially lead
to future concentration and the formation of stars.
\vspace{-4mm}
\subsection{Group Membership}

Since HIJASS J1021+6842 was discovered in a survey of the M\,81 group,
it is natural to assume that it is a member of that group.
Nonetheless, HIJASS J1021+6842 is an unusual system and, before
assuming group membership, it is important to consider evidence in
favor or against this assumption.  An overview of the relative
locations of the M\,81 group members in the plane of the sky is
presented in \citep{Kea02}, their Figure~1.  The systemic velocity of
$\approx$ 30 km\,s$^{-1}$ is certainly consistent with membership in
the M\,81 group \citep{Kea02}.  In fact, it is only offset from M\,81
by 2.3\arcdeg\ in angle (corresponding to a minimum distance of $\sim$
140 kpc) and 40 km\,s$^{-1}$ in radial velocity. In addition, HIJASS
J1021+6842 is situated very close to the M\,81 group member IC\,2574
both in position (projected distance: 110\,kpc) and velocity
($\Delta$v=30\,\kms). This may suggest that IC\,2574 and HIJASS
J1021+6842 are companions, a scenario which is tentatively supported
by the HIJASS observations which indicate a low column density HI
bridge connecting them.

What is the probability that this object is part of the Local Group
rather than the M\,81 group? In that case HIJASS J1021+6842 would
likely be classified as a high velocity cloud (HVC).  There are
several problems with such a hypothesis, however. A strong argument
against it is its fairly high column density, of order $10^{20}$ atoms
cm$^{-2}$ which is at least an order of magnitude higher than what is
found even in the Magellanic Stream \citep{Pea03}.  Given its angular
size, HIJASS J1021+6842 would be classified as a Compact HVC.  Again,
typical column densities for CHVCs are at least an order of magnitude
lower \citep{dHBB02}. An additional strong argument arguing against
HIJASS J1021+6842 being an HVC is that typical HVCs in the direction
of M\,81 have velocities in the range $-$150 to $-$200 km\,s$^{-1}$
\citep{WvW97}, i.e. at much lower velocities.  We therefore conclude
that HIJASS J1021+6842 is indeed associated with the M\,81 group.
\vspace{-2mm}

\section{Conclusions}

We present VLA HI 21cm observations of HIJASS J1021+6842 discovered by
\citet{Bea01} in the direction of the M\,81 group. Given its location,
column densities and systemic velocity, HIJASS J1021+6842 is very
likely a member of the M\,81 group. Perhaps the most striking result
is that the HI emission is distributed over an area of 30\,kpc in
diameter which is much larger than the extent of the largest
dwarf members of the M\,81 group. This, and the lack of detectable
stars makes the HIJASS J1021+6842 system absolutely unique amongst the
M\,81 group members.  Other HI clouds known to date which have no
detected optical counterparts are the SW clump of HI 1225+01
\citep{GH89, Che95, TM97}, HIPASS J0731-69 \citep{Rea01}, and HI
clouds in Virgo \citep{Dea04, Min05}. We detect 1.6 $\times$ 10$^8$
\msun\ of HI distributed over roughly 30 kpc in HIJASS J1021+6842.
Peak HI column densities are of order $1.8 \times 10^{20}$ atoms
cm$^{-2}$, which is well below the empirical threshold for star
formation activity to commence. This may explain why to date no
optical counterpart has been identified.

The individual clouds which make up the system are either self
gravitating entities orbiting within their mutual gravitational
potential, or the densest concentrations of a huge, 30\,kpc diameter,
gently rotating, very low surface density cloud. Assuming that the
entire complex is gravitationally bound, we derive a minimum dynamical
mass of $5.5\times10^9$\,M$_{\odot}$, which would be more than an
order of magnitude more massive than the luminous (HI) mass. It should
be stressed, though, that other scenarios, such as a past stripping
event leading to the formation of tidal debris, may also result in the
observed velocity structure (thus leading to a significantly lower
dynamical mass of the system).

The present observations clearly demonstrate the uniqueness of HIJASS
J1021+6842. Future high sensitivity HI synthesis observations of
HIJASS J1021+6842 and its surroundings and deep optical and UV imaging
will be necessary to elucidate the true nature of this enigmatic
object.

\acknowledgements EDS is grateful for partial support from a NASA
LTSARP grant No.\ NAG5-9221 and the University of Minnesota. We thank
our referee, Dr. Mike Disney, for useful comments which helped to
improve the presentation of this paper.

%\begin{table}
%\dummytable\label{tab:obs}
%this table contains the list of observations
%\end{table}
%
%\begin{table}
%\dummytable\label{tab:results}
%this table contains the imaging parameters
%\end{table}

\clearpage

\clearpage


\begin{thebibliography}{}
\bibitem[Boerngen \& Karachentseva(1982)]{BK82} Boerngen, 
F.~\& Karachentseva, V.~E.\ 1982, Astronomische Nachrichten, 303, 189 
\bibitem[Bournaud et al.(2004)]{Bea04} Bournaud, F., Duc, 
P.-A., Amram, P., Combes, F., \& Gach, J.-L.\ 2004, \aap, 425, 813 
\bibitem[Boyce et al.(2001)]{Bea01} Boyce, P.~J., et al.\ 2001, \apjl,
  560, L127  
\bibitem[Caldwell et al.(1998)]{Cea98} Caldwell, N., Armandroff, T.~E., 
Da Costa, G.~S., \& Seitzer, P.\ 1998, \aj, 115, 535 
\bibitem[Chengalur et al.(1995)]{Che95} Chengalur, J.~N., 
Giovanelli, R., \& Haynes, M.~P.\ 1995, \aj, 109, 2415 
\bibitem[Davies et al.(2004)]{Dea04} Davies, J., et al.\ 2004, \mnras,
349, 922
\bibitem[de Heij, Braun, \& Burton (2002)]{dHBB02} de Heij, V., Braun,
R., \& Burton, W.~B. 2002, \aap, 392, 417 \\
\bibitem[Giovanelli \& Haynes(1989)]{GH89} Giovanelli, R.~\& 
Haynes, M.~P.\ 1989, \apjl, 346, L5
\bibitem[Karachentsev et al.(2002)]{Kea02} Karachentsev, I.~D., et
  al.\ 2002, \aap, 383, 125  
\bibitem[Karachentseva(1968)]{K68}
Karachentseva, V.~E.\ 1968, Commun.\ Byurakan Obs.\ 39, 62 
\bibitem[Karachentseva \& Karachentsev(1998)]{KK98} 
Karachentseva, V.~E.~\& Karachentsev, I.~D.\ 1998, \aaps, 127, 409  
\bibitem[Kennicutt(1989)]{K89} Kennicutt, R.~C.\ 1989, \apj, 344, 685
\bibitem[Makarova et al.(2002)]{Mea02} Makarova, L.~N., et al.\ 2002,
  \aap, 396, 473  
\bibitem[Mateo(1998)]{M98} Mateo, M. L. 1998, \araa, 36, 435
\bibitem[Minchin et al.(2005)]{Min05} Minchin, R., et al.\ 
2005, \apjl, 622, L21
\bibitem[Ott et al.(2001)]{Oea01} Ott, J., Walter, F., 
Brinks, E., Van Dyk, S.~D., Dirsch, B., \& Klein, U.\ 2001, \aj, 122, 3070 
\bibitem[Puche et al.(1992)]{PWBR92} 
Puche, D., Westpfahl, D., Brinks, E., \& Roy, J.\ 1992, \aj, 103, 1841 
\bibitem[Putman et al.(2003)]{Pea03} Putman, M.~E., Staveley--Smith, L.,
Freeman, K.~C., Gibson, B.~K., \& Barnes, D.~G. 2003, \apj, 586, 170
\bibitem[Ryder et al.(2001)]{Rea01} Ryder, S.~D., et al.\ 
2001, \apj, 555, 232 
\bibitem[Skillman(1996)]{S96} Skillman, E.~D.\ 1996, ASP 
Conf.~Ser.~106: The Minnesota Lectures on Extragalactic Neutral Hydrogen, 208 
\bibitem[Turner \& MacFadyen(1997)]{TM97} Turner, 
N.~J.~J.~\& MacFadyen, A.\ 1997, \mnras, 285, 125
\bibitem[van den Bergh(1959)]{vdb59} van den Bergh, S.\ 1959,
Publ.\ of D.D.O., Vol.II, No 5, 147 
\bibitem[Wakker \& van Woerden (1997)]{WvW97} Wakker, B.~P., \& van
Woerden, H. 1997, \araa, 35, 217
\bibitem[Walter \& Brinks(1999)]{WB99} Walter, F.~\& Brinks, E.\ 1999,
  \aj, 118, 273  
\bibitem[Walter et al.(2002)]{WWMS02} 
Walter, F., Weiss, A., Martin, C., \& Scoville, N.\ 2002, \aj, 123, 225 
\bibitem[Yun, Ho, \& Lo(1994)]{YHL94} Yun, M.~S., Ho, P.~T.~P., \& Lo, K.~Y.\ 1994, \nat, 372, 530 
\end{thebibliography}
\end{document}